# Test of a Multilayer Dose-Verification Gaseous Detector with Raster Scan Mode Proton Beams

Kyong Sei Lee*†, Byungsik Hong, Minho Kang, Seunkyung Lee, and Sung Keun Park
*Korea University, Seoul 136-713*

Sung Hwan Ahn*, Youngyih Han
*Department of Radiation Oncology, Samsung Medical Center, Seoul 135-710*

Sang Yeol Kim
*NoticeKorea, Anyang 431-815*

A multilayer gaseous detector has been developed for the fast dose-verification measurements of raster-scan-mode therapeutic beams in particle therapy. The detector, which was constructed with eight thin parallel plate ionization chambers (PPICs) and polymethyl methacrylate (PMMA) absorber plates, is closely tissue-equivalent in a beam's eye view (BEV). The gas-electron signals, collected on the strips and pad arrays of each PPIC, were amplified and processed with a continuous charge-integration mode. The detector was test with 190-MeV raster-scan-mode beams that were provided by the Proton Therapy Facility at Samsung Medical Center, Seoul, South Korea. The detector responses, of the PPICs for a 190-MeV raster-scan-mode proton beam, agreed well with the doses predicted by the GEANT4 simulations. Furthermore, in this study it was confirmed that the detector simultaneously tracked the doses induced at the PPICs by the fast-oscillating beam, with a scanning speed of 2 m s$^{-1}$. Thus, it is anticipates that the present detector, composed of thin PPICs and operating in the charge-integration mode, will allow medical scientists to perform reliable fast dose verification measurements for the typical dynamic mode therapeutic beams.

PACS numbers: 29.40.Cs, 29.27.Fh, 87.52.-g.

Keywords: Particle therapy, Parallel plate ionization chambers, Dose verification, Raster-scan-mode proton beam.

## I. INTRODUCTION

In radiation therapy, proton [1-5] and heavy-ion beams [6-10] have a great advantage over X-rays because of the high radiation biological effectiveness (RBE) of providing biological damage on malignant tumors and also of suppressing the fast recovery of the DNA. Furthermore, the well-controlled delivery, of the radiation to the planned treatment volume (PTV) and the well-defined depth of interaction, minimize an unnecessary delivery of the dose to normal tissue or sensitive organs surrounding the cancer tumor. As the technology of particle therapy becomes more sophisticated, it is necessary that there is a more accurate and swifter confirmation of the proper measurement and the verification of the doses to be delivered to patients.

In heavy-ion therapy, the wire chambers operating in an ionization-chamber mode have been typically used for monitoring beams and precise dose-verification measurements [9]. The detector responses, which have to be interpreted for the dose rate and to confirm the beam, are measured in the two coordinates vertical to the beam direction and as functions of the depth of interaction in a water phantom, by consecutively changing the thickness of the water lying in front of the detector surface.

However, simultaneous measurement over the whole depth range will be fairly conducive to reducing the scan time and labor required for the dose verification process. Furthermore, tracking the movement of the fast oscillating raster-scan or spot-scan beams [6, 7] at each depth position of the PTV would also be advantageous to protecting the normal tissue surrounding the cancer tumor, and thereby reduce undesirable damage that can be caused by the lateral penumbra of the beam. The gaseous detector for the dose verification measurement of particle therapy, which was developed in the present research, is briefly summarized as follows:

(1) The detector is composed of eight thin and uniform PPICs whose signals are collected by fine strips and pad arrays with a pitch of 1.25 mm.

(2) The mean material density of the detector that the therapeutic beams experienced was closely adjusted to be tissue equivalent (1.003 g cm$^{-3}$) [11] in a beam's eye view within an accuracy of about 0.15% (standard deviation).

(3) The detector response data are processed by amplifying and integrating the gas-electron signals over continuous system clock-time intervals of 28.6 μs.

(4) The detector response data are reconstructed every 228.6 μs, which allows for a reproduction of a maximum of 8×4372 frames per second of dose images.

In this study, Section 2 is a description of the structure of the PPIC and the construction of the detector composed of eight PPICs, each equipped with a 256-channel signal-processing and data-transfer electronics. Section 3 is a report of the test results performed with 190-MeV raster-scan-mode beams provided by a proton cyclotron at the Samsung Medical Center, Seoul, South Korea. Lastly, in Section 4, there is a conclusion about the present research and a discussion about the future tasks in research and development for the practical applications.

## II. CONSTRUCTION OF THE DETECTOR

The multilayer gaseous detector in the present research and development was designed for the dose verification measurement of therapeutic proton beams with a maximum energy of 200 MeV, whose predicted distal edge in the detector was 27 g·cm$^{-2}$. The detector was assembled with interlaced eight-layer thin PPICs and absorbers made of PMMA plates (see Fig. 1). The mean density of the detector that therapeutic particles experience was closely adjusted to 1.003 g cm$^{-3}$ by properly placing the PMMA plates with thicknesses of 5 and 10 mm. The size of the gap lying between two adjacent PPICs, allowing the placing the PMMA plates, was adjusted to 3.6 g·cm$^{-2}$.

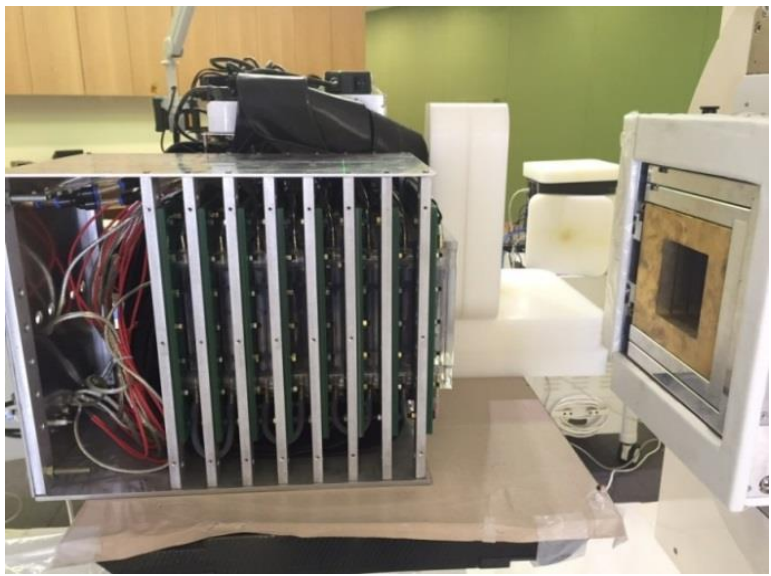

Fig. 1. Gaseous detector constructed interlaced eight-layer thin PPICs and absorbers made of PMMA plates.

To achieve a complete three-dimensional measurement with high voxel granularity, additional PMMA absorber plates can be placed anterior to the front surface of the detector. For example, the scan measurements for the designed beam can be repeated with absorber thicknesses increasing from 0 to 3.2 g·cm$^{-2}$, in 0.4-g·cm$^{-2}$ steps. Then, there is a reconstruction of a complete set of data with the desired voxel granularity, with a total of 72 depth steps in the beam direction. The amount of time required to complete a single dose-verification measurement with a 10-s scan measurement for each step was expected to be less than 300 s. The single 10-s beam to be measured at each step and may contain many beam components with different energies, which are required to properly cover the tumor size.

The structure of a PPIC, designed in the present research, is illustrated with a schematic diagram in the top figure of Fig. 2. The thickness of the induction region, formed between the cathode and the signal planes, was adjusted to 3.8 mm. The copper patterns for the cathode and signal planes were printed on 1.6-mm thick printed circuit boards (PCBs). The signal plane is composed of 1.25-mm-pitch strips and pad arrays, printed on the top layer of the PCB, as can be seen in the bottom figure of Fig. 2. The strips and the pad arrays, 128 of each, were assigned to measure the separate detector responses in the vertical ($y$) and horizontal ($x$) directions, respectively. The 100-µm-wide traces, required for the electrical connections for the pad arrays, were printed on the bottom layer.

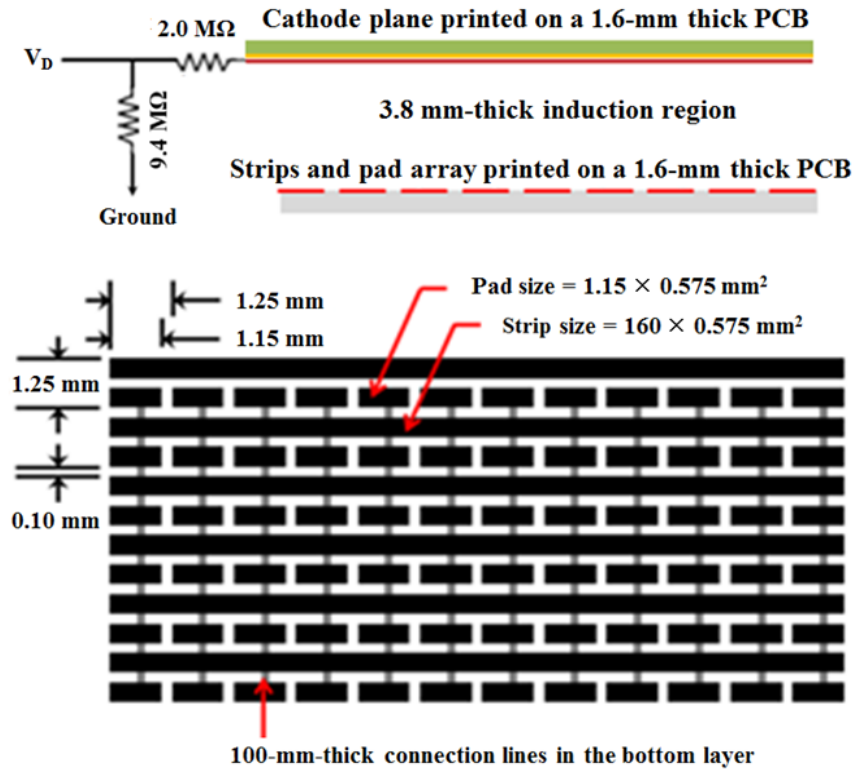

Fig. 2. Copper pattern for the signal plane composed of 1.25-mm-pitch strips and pad arrays printed on the top layer of a 1.6-mm-thick PCB. The 100-µm-wide traces required for the electrical connections for the pads were printed on the bottom layer of the PCB.

The typical electrical potential of −570 V, applied to the cathode plane, provides a uniform electric field intensity of 1.5 kV cm$^{-1}$ in the induction volume of the PPIC. It was shown in previous research [12, 13] that the amplification of the gas electrons was fairly insignificant, over the wide range of electric-field intensity, when the present ionization-mode detector was operated with a gas mixture of 70% Ar + 30% $CO_2$.

The 256-channel electronics board for the signal process and the data transfer, each assigned to a PPIC, was composed of four 64-channel charge integrators, eight amplifiers, an eight-channel analog-digital-converter (ADC) processor, a field-programmable-gate-array (FPGA) digital processor, and a USB3 interface processor. The details for this electronics design are described in a previous report [13-15].

The gas electron signals, collected in the strips and pad arrays of the detectors, were integrated every 28.6 µs and converted to the integrated charge values (channel responses). The maximum sensitivity of the proton-induced current per channel was adjusted to 20 nA.

### III. TESTS OF THE DETECTOR WITH RASTER SCAN MODE PROTON BEAMS

The performance of the present dose-verification detector was examined with 190-MeV raster-scan-mode beams that were provided by the Proton Cyclotron Facility at Samsung Medical Center, Seoul, South Korea. The detector was installed at a distance of 10 cm from the rectangular copper collimator mounted in the rotating beam gentry, where a compensation Bolus, which is required to further focus a dose to a patient's tumor, is to be placed (see Fig. 3).

A schematic of a typical raster scan, with a thin pencil-shape pristine proton beam, can be seen in Fig. 4. The blue line indicates the passage of the Bragg peak of a monoenergetic beam on the surface, lying at a depth in the PTV. The depth of a scan in the PTV can be adjusted by changing the energy of the beam.

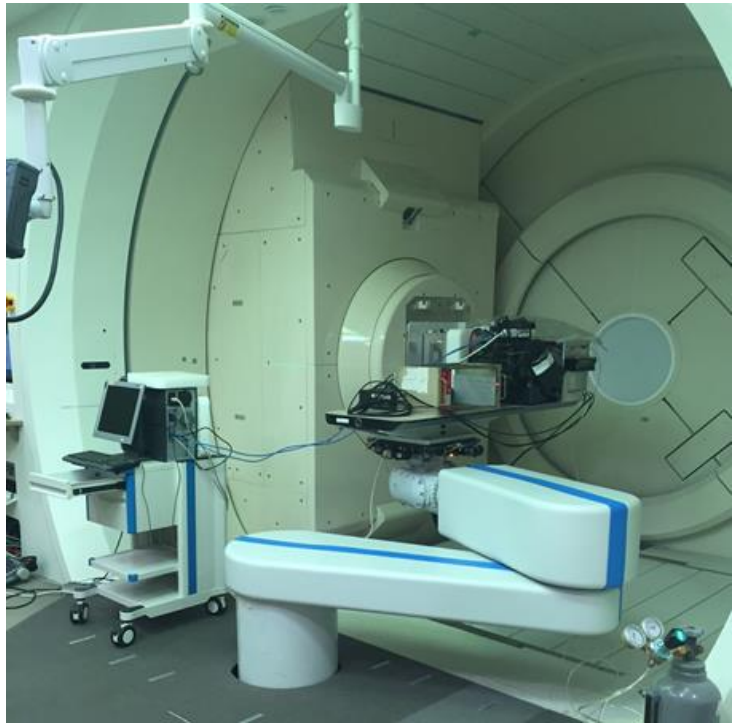

Fig. 3. Detector installed at a distance of 10 cm from the rectangular copper collimator, mounted in the rotating beam gentry, where a compensation Bolus, required to further focus the dose to the patient's tumor, is to be placed.

In Fig. 5, the time responses are shown for the first PPIC for a 1.0-nA 190-MeV raster-scan-mode proton beam with a scanning speed of 10 m s$^{-1}$, measured by the pad arrays ($x$) and the strips ($y$) for a 1-s period (*ibid.*, the top and the bottom panels, respectively). The field size of the beam was adjusted to $11 \times 11$ cm$^2$. The substructures of the time profiles in Fig. 5 respect the movement of the beam in the detector, equivalent to a PTV, as illustrated in Fig. 4. The required duration for the completion of a single raster scan was measured as 233 ms.

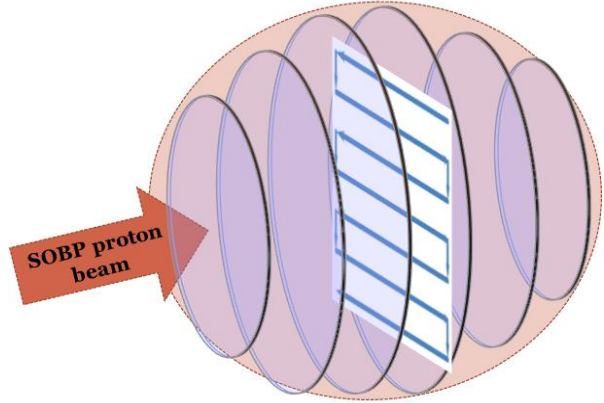

Fig. 4. Typical raster scan with thin pencil-shape pristine proton beams. The blue line indicates the passage of the Bragg peak of a monoenergetic beam on the surface lying at a depth in the PTV.

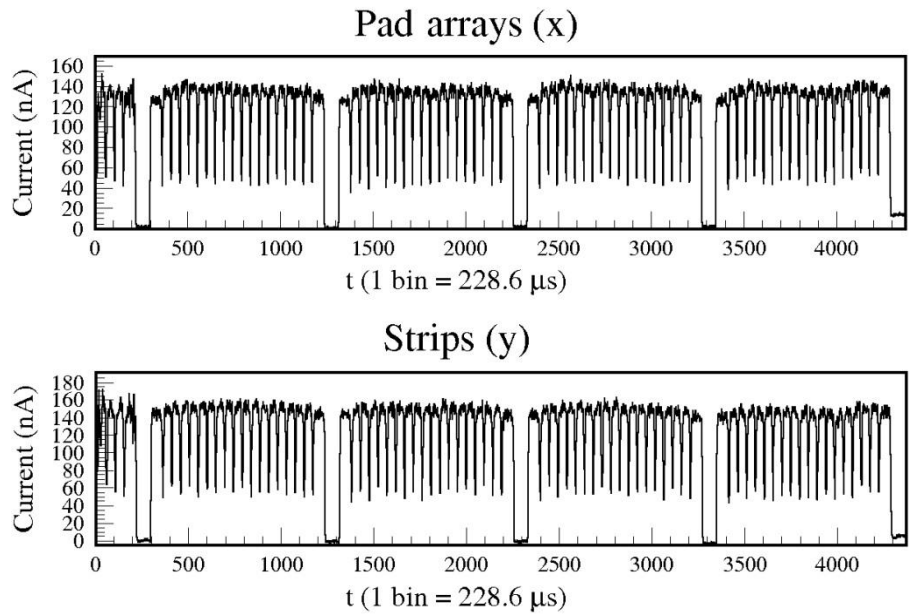

Fig. 5. Time responses of the first PPIC to the raster-scan mode beam for 1 s measured by pad arrays (*x*) at the top and strips (*y*) at the bottom panels, respectively.

The procedure of the fast oscillating raster-scan-mode beam, which induces the dose at each depth in the detector phantom, can be visualized in pseudo-two-dimensional (2D) images where the data, obtained by multiplying the channel-response functions in the x and the y directions (*i.e.*, $F(x,y) = \sqrt{f(x) \times f(y)}$) are sequentially added. Figure 6 shows the pseudo-2D images for the raster scan mode beam, as measured at the first PPIC at 0.014 (top left), 0.048 (top middle), 0.128 (top right), 0.256 (bottom left), 1 s (bottom middle), and 10 s (bottom right), after the start of the measurement. The width (full width at half maximum) of the raster scan mode beam was measured as about 14 mm (see Fig. 6). As shown in the bottom-right pseudo-2D image in Fig. 6, the flatness of the proton field, measured for 10 s, is fairly satisfactory for achieving the accuracy required for the precision radiological operation. The movement of the beam in the raster-scan motion, measured at the first PPIC for first 48 ms, can be seen in the pseudo-2D images (Fig. 7). They are fairly conducive to the precision space-time analysis and the relevant visualization for the rapidly oscillating dynamic mode beam.

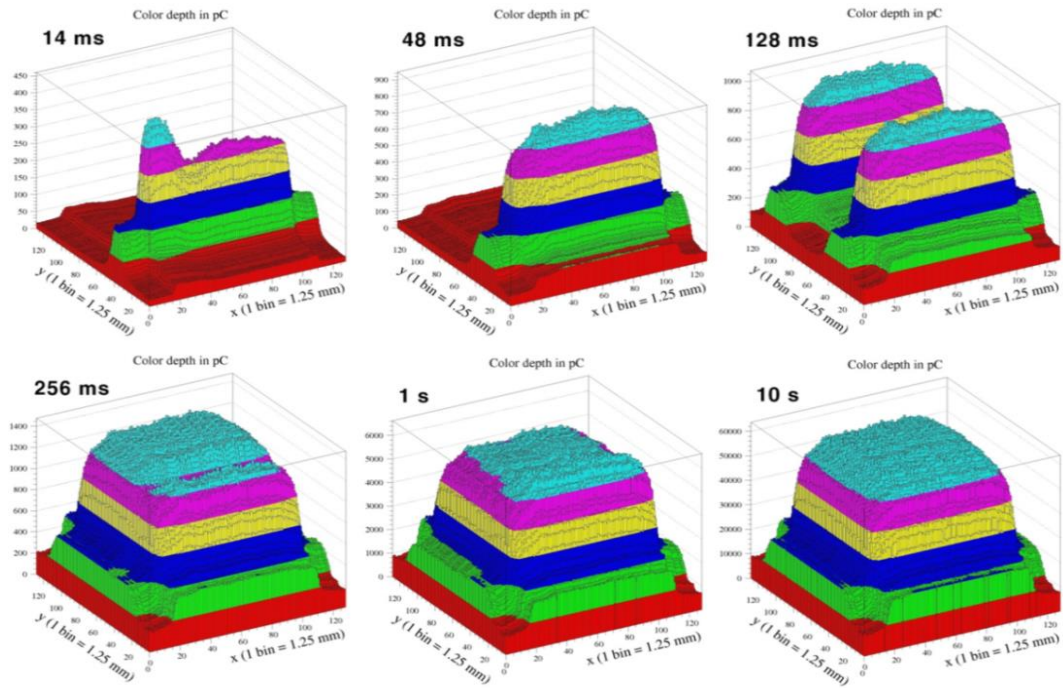

Fig. 6. Pseudo 2D images for the raster-scan-mode beam measured at the first PPIC at 0.014 (top left), 0.048 (top middle), 0.128 (top right), 0.256 (bottom left), 1 s (bottom middle), and 10 s (bottom right) after the start of the measurement.

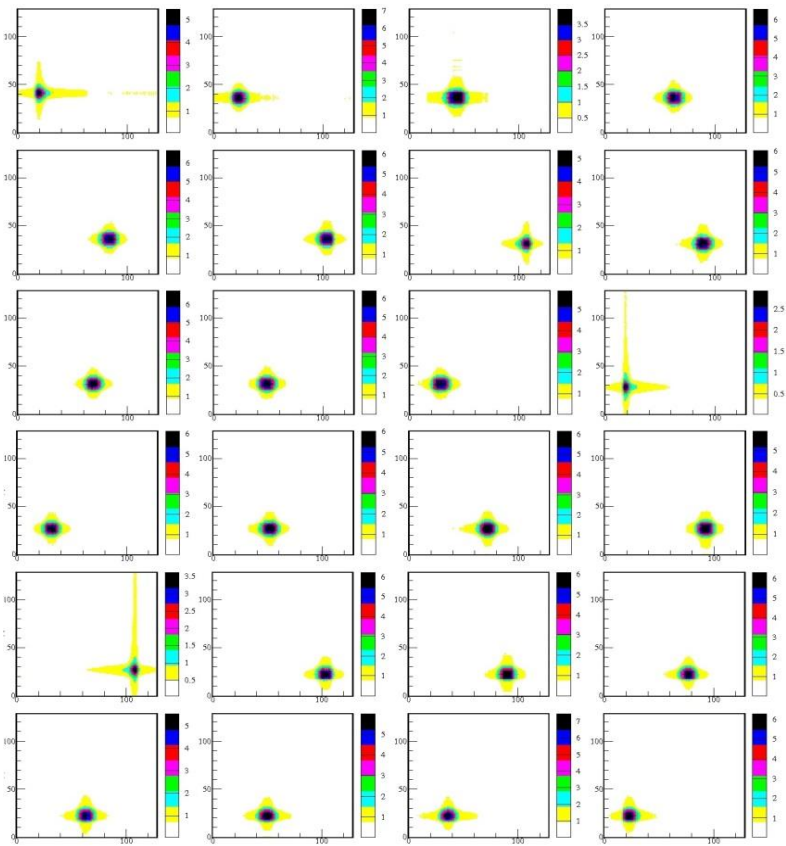

Fig. 7. Movement of the beam in the raster-scan motion measured by the first PPIC for first 48 ms.

Figure 8 shows total detector responses (full circles) of the PPICs denoted by "$D_i$," with $i$ ranging from 1 to 7, and installed at the first seven depths in the tissue equivalent detector of a 1.6-nA 190-MeV raster-scan mode proton beam, with a duration of 2.4 s and with a scanning speed of 2 m s$^{-1}$. The field size of the beam was adjusted to $10 \times 10$ cm$^2$. The solid curve in Fig. 8 is the one-dimensional Bragg curve that was predicted by performing a GEANT4 simulation [16-19]. The positions of the Bragg peak and the distal edge of the 190-MeV protons are expected to be 23.3 and 24.0 g cm$^{-2}$, respectively. For the proper comparison of the data with the Bragg curve obtained by the GEANT4 simulation, the relative strengths of the PPIC responses were adjusted by normalizing the first detector response (denoted by "$D_1$") to the simulated dose rate at the same depth. Figure 9 shows a two-dimensional representation of the dose simulated by the GEANT4 program, as a function of the depth and one direction perpendicular to the depth direction. The width of the area of the 190-MeV raster-scan mode beam becomes broader by about 15% as the beam reaches the Bragg peak.

The distributions, of the detector responses of the eight PPICs in the $x$ and the $y$ directions for the 1.6-nA 190 MeV raster-scan-mode proton beam, are shown in the first and second columns in Fig. 10, respectively. The relatively large statistical fluctuations of the distributions were attributed to the absence of proper calibrations for the channel sensitivities of the electronics, which will be properly performed using a raster scan mode beam whose field area is fairly flat and wider than the detector active area. Nevertheless, the detector responses measured at the first seven depths, as shown in Figs. 8 and 10, are well coincided with the simulated dose rates.

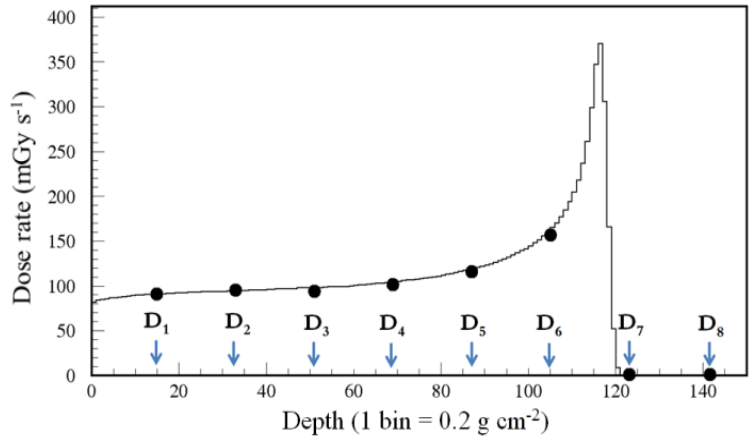

Fig. 8. Total detector responses (full circles) of the PPICs, denoted by "$D_i$," with $i$ ranging from 1 to 7, and installed at the first seven depths in the tissue-equivalent detector to a 1.6-nA 190-MeV raster-scan-mode proton beam for 2.4 s and with a scanning speed of 2 m s$^{-1}$.

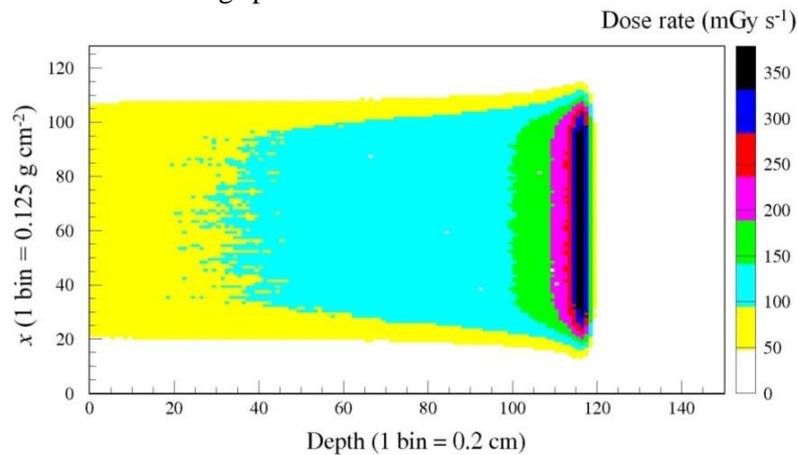

Fig. 9. Two-dimensional representation of the dose simulated by the GEANT4 program as a function of the depth and one direction perpendicular to the depth direction.

The figures, in the third column of Fig. 10, show pseudo-2D images of the detector responses, measured for the 1.6-nA 190-MeV raster-scan-mode proton beam at the first seven depths. The pseudo 2D-images visualize that the area of the radiation field and the strength increases as the depth increases (see Fig. 10). However, it should be noted that the strength near the periphery of the radiation field (halo) is inevitably exaggerated, as shown in the pseudo-2D image that has been reconstructed by the multiplication of the two response functions. It is not equivalent to a 'true' 2D image, unless the raster scan is performed with a beam with an infinitely narrow width. Thus, the use of the pseudo-2D-images should be restricted to the confirmation of an area and the flatness of a radiation field, as well as verification of the movement of the beam in the detector.

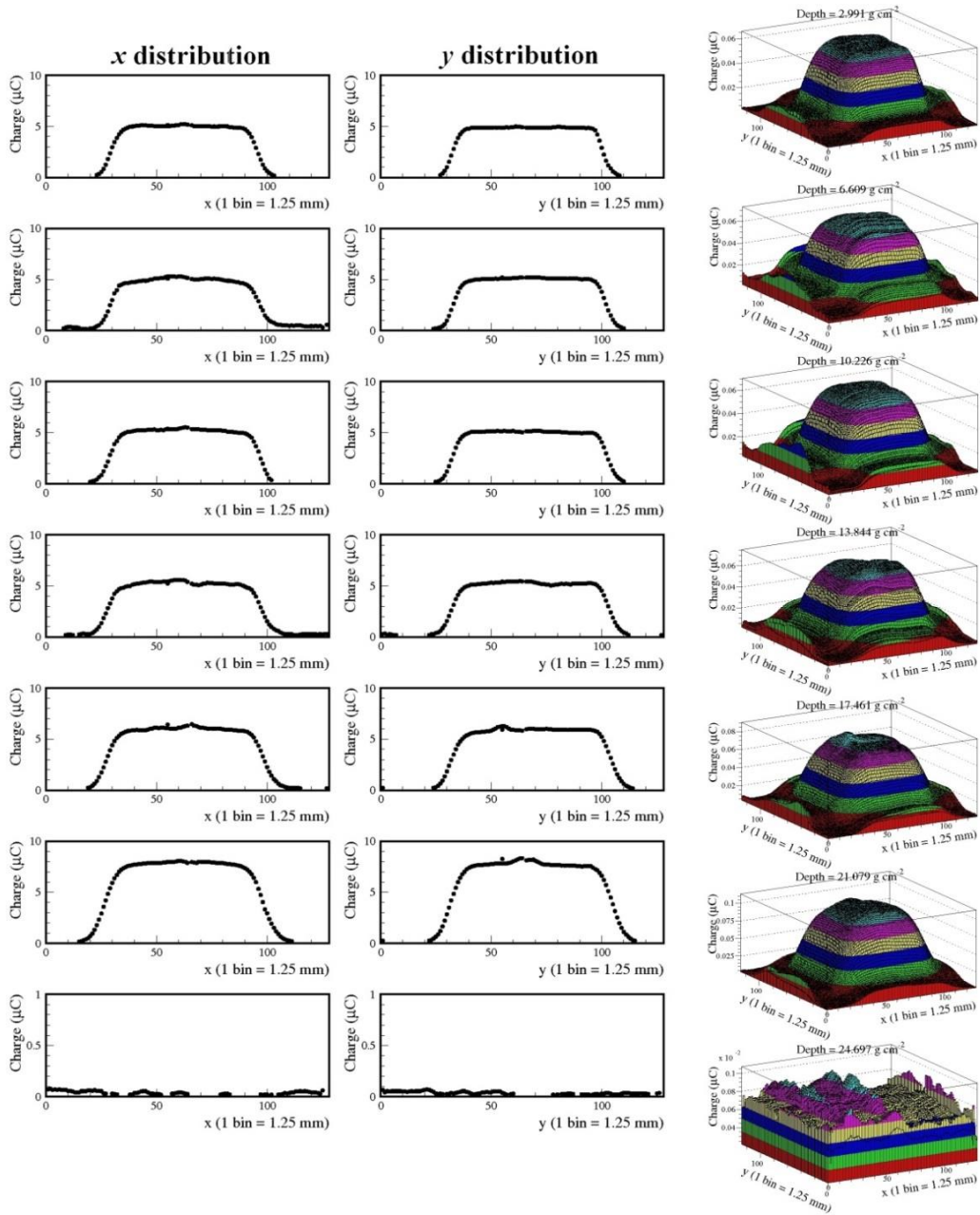

Fig. 10. Distributions of the detector responses of the first seven PPICs in the *x* (first column) and the *y* (second column) directions for the 1.6-nA 190-MeV raster-scan-mode proton beam. The corresponding pseudo-2D images are shown in the third column.

## IV. Conclusion

In this study we have constructed a gaseous detector composed of multilayer PPICs and PMMA absorbers, and examined a proposed radiation technique for fast dose-verification measurements in particle therapy, with the use of therapeutic proton beams provided by the Proton Therapy Facility at Samsung Medical Center, Seoul, South Korea. The prototype detector was manufactured to be closely tissue equivalent in a beam's eye view. The performance of the prototype detector was confirmed using raster scan mode proton beams, whose energy, field area, scan speed, and pattern are typically used in the dynamic mode treatments. The conclusions of the present research, performed with the prototype detector, are summarized as follows:

(1) The flatness of a radiation field of a raster-scan-mode proton beam, provided by the Proton Therapy Facility (Samsung Medical Center Seoul, South Korea) is fairly satisfactory for achieving the accuracy required for a precision radiological operation in particle therapy.

(2) The movements of the fast raster-scanning proton beam in the detector system were well reproduced in the pseudo-2-D images that were reconstructed from the data and time responses. They will be fairly conducive to a precision analysis and a proper visualization in clinical procedure with dynamic-mode therapeutic beams.

(3) The detector responses of the PPICs, placed in the detector, are well coincided with the dose rates predicted by the GEANT4 code.

The development of detailed clinical procedures for dose verification is a future research task with the present detector. It includes proper procedures for the calibration of channel sensitivities and for multi-step measurements to obtain the voxel data that completely covers fine depth positions, as well as a critical evaluation of the performance of the detector when applied to the actual radiation treatment process. A promising new development for the detector would entail using the current research and development results of heavy-ion therapy, after construction of the first heavy ion medical facility is completed in the near future in South Korea.

## Acknowledgment

This study was supported by the National Research Foundation of Korea (Grant Numbers NRF-2013R1A1A2060257 and NRF-2013M2B2A9A-03050128).